\documentclass[12pt,leqno]{article}
\title{Critical mass of bacterial populations and critical temperature of self-gravitating Brownian particles
 in two dimensions}

\usepackage{amsthm,amsmath,amssymb,a4wide,psfig,epsf}
\def\mb#1{\setbox0=\hbox{$#1$}\kern-.025em\copy0\kern-\wd0
\kern-0.05em\copy0\kern-\wd0\kern-.025em\raise.0233em\box0}

\begin{document}

\author{Pierre-Henri Chavanis}
\maketitle
\begin{center}
Laboratoire de Physique Th\'eorique (UMR 5152 du CNRS), Universit\'e Paul Sabatier,\\
118, route de Narbonne, 31062 Toulouse Cedex, France\\ E-mail:
{\it chavanis{@}irsamc.ups-tlse.fr}

%\date{}
\vspace{0.5cm}
\end{center}

\begin{abstract}
We show that the critical mass $M_c=8\pi$ of bacterial populations in
two dimensions in the chemotactic problem is the counterpart of the
critical temperature $T_c=GMm/4k_{B}$ of self-gravitating Brownian
particles in two-dimensional gravity. We obtain these critical values
by using the Virial theorem or by considering stationary solutions of
the Keller-Segel model and Smoluchowski-Poisson system. We also
consider the case of one dimensional systems and develop the
connection with the Burgers equation. Finally, we discuss the
evolution of the system as a function of $M$ or $T$ in bounded and
unbounded domains in dimensions $d=1$, $2$ and $3$ and show the
specificities of each dimension. This paper aims to point out the
numerous analogies between bacterial populations, self-gravitating
Brownian particles and, occasionally, two-dimensional vortices.

\vskip0.5cm {\it Keywords:} Chemotaxis, two-dimensional gravity, self-gravitating Brownian particles, nonlinear meanfield Fokker-Planck equations, Burgers equation, two-dimensional turbulence
\end{abstract}

% main text

\section{Introduction}
\label{sec_intro}

In many fields of physics, astrophysics and biology, one is
confronted with the description of the evolution of a system of
particles which self-consistently attract each other over large
distances. One difficulty and richness of the problem arises from
the long-range nature of the potential of interaction
\cite{houches}. This is the case, for example, in biology in
relation with the process of chemotaxis \cite{murray}. The
chemotactic aggregation of bacterial populations (like {\it
Escherichia coli}) or amoebae (like {\it Dictyostelium discoideum})
is usually studied in the framework of the Keller-Segel model
\cite{keller} which describes the collective motion of organisms that
are attracted by a chemical substance (pheromone) that they produce
themselves.  The Keller-Segel (KS) model involves a drift-diffusion
equation describing the evolution of the concentration of the bacteria
in the gradient of concentration of the secreted chemical. In
the simplest formulation, the concentration of the chemical is related
to the concentration of bacteria by a Poisson equation (this is valid
in a limit of large diffusivity of the chemical and for sufficiently
large concentrations)
\cite{jager}. These equations have been studied by applied
mathematicians who obtained rigorous results for the existence and
unicity of the solutions and for the conditions of blow-up, modeling
chemotactic collapse, in different dimensions of space
\cite{horstmann}. In particular, in $d=2$, there exists a critical
mass $M_c$ (independent on the size of the domain) above which the
system collapses and forms a Dirac peak. Alternatively, for $M<M_{c}$,
the system spreads to infinity in an unbounded domain or tends to a
stationary state in a bounded domain.

Gravity is another example of long-range attractive potential of
interaction. In a series of papers, Chavanis \& Sire
\cite{crs,sc,post,tcoll,banach,crrs,sopik,virial1,virial2} have studied a
model of self-gravitating Brownian particles in various dimensions of
space. In statistical mechanics, this model is associated with the
canonical ensemble in which the temperature is fixed. A lot of
analytical results have been obtained and an almost complete
description of the system, for all the phases of the dynamics
(pre-collapse and post-collapse), has been given in the overdamped
limit of the model. In that limit, the evolution of the density of the
self-gravitating Brownian gas is governed by the Smoluchowski-Poisson
(SP) system. The Smoluchowski equation is a drift-diffusion equation
of a Fokker-Planck type. For self-gravitating particles, the
gravitational potential inducing the drift is produced by the density
of particles through the Newton-Poisson equation. It turns out
that the Smoluchowski-Poisson system is isomorphic to the simplified
version of the Keller-Segel model of chemotaxis provided that the
parameters are suitably re-interpreted, as discussed in
\cite{crrs}. In particular, for the 2D self-gravitating Brownian gas,
there exists a critical temperature $T_c$ (independent on the size of
the domain) below which the system collapses and forms a Dirac peak.
This is the counterpart of the critical mass of bacterial populations.
For $T>T_{c}$, the system evaporates in an unbounded domain or tends
to a statistical equilibrium state in a bounded domain.

The object of this paper is to emphasize the parallel between these
two systems. In Secs. \ref{sec_ks} and \ref{sec_sp}, we use the Virial
theorem to derive the critical mass of bacterial populations and the
critical temperature of self-gravitating Brownian particles in two
dimensions.  In Sec. \ref{sec_ep}, we show that these critical values
can also be obtained by considering stationary solutions of the
Keller-Segel model and Smoluchowski-Poisson system. In Sec. \ref{sec_ab}, we consider the one dimensional problem and  point out the connection between the
Smoluchowski-Poisson system (or the Keller-Segel model) and the
Burgers equation. We use this analogy to provide the general solution
of these equations in $d=1$ in bounded and unbounded domains. Finally, in Sec. \ref{sec_dyn}, we
provide a summary of the results obtained by Chavanis \& Sire
\cite{crs,sc,post,tcoll,banach,crrs,sopik,virial1,virial2} for self-gravitating
Brownian particles and adapt them to the context of chemotaxis to
clearly show the link between these two problems.

\section{The Keller-Segel model}
\label{sec_ks}

The dynamical evolution of biological populations like bacteria,
amoebae, cells... that are attracted by a substance that they emit
themselves, is often described by the Keller-Segel (KS) model
\cite{keller}. In its simplest form, it can be written as
\begin{equation}
\label{ks1} {\partial\rho\over\partial t}=D\Delta\rho-\chi\nabla \cdot (\rho\nabla c),
\end{equation}
\begin{equation}
\label{ks2}\Delta c=-\lambda\rho.
\end{equation}
Equation (\ref{ks1}) is a drift-diffusion equation for the cell
density $\rho({\bf r},t)$. The diffusion term takes into account the
erratic motion of the cells (like in Brownian theory) and the drift
term with $\chi>0$ takes into account the chemotactic attraction  (one
could also consider the case $\chi<0$ where the secreted substance is
a noxious substance, like a poison, so that chemotaxis is
repulsive). It is directed along the gradient of concentration $c({\bf
r},t)$ of the secreted chemical. In the simplest formulation \cite{jager}, the
production of the chemical by the cells is described by a Poisson
equation (\ref{ks2}). This is valid in a limit of high diffusivity of
the chemical and for sufficiently large concentrations (see Appendix
\ref{sec_bc}). The Keller-Segel model (\ref{ks1})-(\ref{ks2}) must be
supplemented by appropriate boundary conditions. A first physical
boundary condition is that the current ${\bf
J}=D\nabla\rho-\chi\rho\nabla c$ is parallel to the boundary of the
domain (or vanishes at infinity in an unbounded domain) so that the
total mass is conserved.  On the other hand, when dealing with the
reduced Keller-Segel model (\ref{ks1})-(\ref{ks2}), one usually
assumes that $c({\bf r},t)$ is the solution of the Poisson equation
(\ref{ks2}) in an infinite domain with the usual gauge condition. This
yields $c({\bf r},t)={\lambda\over (d-2)S_{d}}\int\rho({\bf
r}',t)|{\bf r}-{\bf r}'|^{-(d-2)}d{\bf r}'$ (in $d\neq 2$) or $c({\bf
r},t)={\lambda\over 2\pi}\int\rho({\bf r}',t)\ln|{\bf r}-{\bf
r}'|d{\bf r}'$ (in $d=2$) whenever the domain containing the bacteria
is finite (box) or infinite (here $S_{d}$ denotes the surface of a
unit sphere in $d$-dimensions). A physical discussion of the boundary
conditions and of the limitations of the reduced Keller-Segel model
(\ref{ks1})-(\ref{ks2}) is provided in Appendix \ref{sec_bc}.

We now proceed in deriving an exact relation that is similar to the
Virial theorem in astrophysics (see Sec. \ref{sec_sp}). As discussed
in \cite{virial1}, it is convenient to take the origin of the system
of coordinates at the center of mass which is a fixed
quantity. Multiplying Eq.~(\ref{ks1}) by $x_{i}x_{j}$ and integrating
over the entire domain, we get
\begin{equation}
\int {\partial\rho\over\partial t} x_{i}x_{j}\,d{\bf r}=\int
x_{i}x_{j} {\partial\over\partial x_{k}}
\biggl (D{\partial \rho\over\partial x_{k}}-\chi\rho
{\partial c\over\partial x_{k}}\biggr ) \,d{\bf r}.
\label{ks3}
\end{equation}
Introducing the ``tensor of inertia''
\begin{equation}
I_{ij}=\int\rho x_{i}x_{j}\,d{\bf r},\label{ks4}
\end{equation}
and integrating the second term by parts twice, we find that
\begin{equation}
{1\over 2}{dI_{ij}\over dt}=DM\delta_{ij}+\chi W_{ij},\label{ks5}
\end{equation}
where $M=\int \rho d{\bf r}$ is the total mass of cells and $W_{ij}=W_{ji}$ is the ``potential energy tensor'' \cite{virial1}:
\begin{equation}
W_{ij}=\int\rho\ x_{i}{\partial c\over\partial x_{j}}\,d{\bf
r}.\label{ks6}
\end{equation}
If the system is confined within a box, we have to account for
boundary terms. The first integration by parts in Eq.~(\ref{ks3})  yields
a residual term
\begin{equation}
\int \nabla \cdot \lbrack x_{i}x_{j}(D\nabla
\rho-\chi\rho\nabla c)\rbrack \,d{\bf r}=\oint x_{i}x_{j}(D\nabla
\rho-\chi\rho\nabla c)\cdot d{\bf S},\label{ks7}
\end{equation}
where $d{\bf S}$ is the surface element normal to the frontier of the
confining box. By virtue of the conservation of mass, the diffusion
current in Eq.~(\ref{ks1}) is always perpendicular to the surface
vector and consequently the term (\ref{ks7}) vanishes.  The second
integration by parts yields
\begin{equation}
-\int {\partial\over\partial x_{k}}\biggl\lbrack
D\rho (x_{j}\delta_{ki}+x_{i}\delta_{kj}) \biggr\rbrack\,d{\bf
r}=-\oint
D\rho(x_{j}\delta_{ki}+x_{i}\delta_{kj})\,dS_{k}.\label{ks8}
\end{equation}
Therefore, the general form of the Virial theorem for the chemotactic
problem, taking into account boundary terms, is
\begin{equation}
{1\over 2}{dI_{ij}\over dt}=DM\delta_{ij}+\chi W_{ij}-{1\over 2}\oint
D\rho (x_{j}\,dS_{i}+x_{i}\,dS_{j}).\label{ks9}
\end{equation}
By contracting the indices, we get the scalar Virial theorem
\begin{equation}
{1\over 2}{dI\over dt}=dDM+\chi W_{ii}-D\oint \rho\,{\bf r}\cdot d{\bf
S},\label{ks10}
\end{equation}
where
\begin{equation}
I=\int \rho r^{2}\,d{\bf r},\label{ks11}
\end{equation}
is the moment of inertia and
\begin{equation}
W_{ii}=\int \rho {\bf r}\cdot \nabla c \, d{\bf r},\label{ks12}
\end{equation}
is the Virial. If the density $\rho_{b}$ is uniform on the
edge of the box (this is the case at least for a spherically
symmetric system), we get
\begin{equation}
\oint \rho\,{\bf r}\cdot d{\bf S}=\rho_{b}\oint {\bf r}\cdot d{\bf S}=\rho_{b}\int \nabla\cdot {\bf r} \, d{\bf r}=d\rho_{b}V,
\label{ks13}
\end{equation}
where $V$ is the  volume of the confining box. Thus,
\begin{equation}
{1\over 2}{dI\over dt}=dDM+\chi W_{ii}-dD\rho_{b}V.\label{ks14}
\end{equation}
On the other hand, adapting the results of the Appendix of Ref
\cite{virial1} to the present situation, we find for $d\neq 2$ that
$W_{ii}=(d-2)W$ where $W=-{1\over 2}\int \rho c d{\bf r}$ is the
``potential energy''. In that case, we get
\begin{equation}
{1\over 2}{dI\over dt}=dDM+(d-2)\chi W-D\oint \rho\,{\bf r}\cdot d{\bf
S}, \qquad (d\neq 2).\label{ks15}
\end{equation}
For $d=2$, we find instead that $W_{ii}=-\lambda
M^{2}/(4\pi)$  and we obtain
\begin{equation}
{1\over 2}{dI\over dt}=2DM-{\lambda \chi M^{2}\over 4\pi}-D\oint \rho\,{\bf r}\cdot d{\bf
S} \qquad (d= 2).\label{ks16}
\end{equation}
At equilibrium ($\dot I=0$), the Virial theorem (\ref{ks10}) reduces to
\begin{equation}
dDM+\chi W_{ii}=D\oint \rho\,{\bf r}\cdot d{\bf
S}.\label{ks17}
\end{equation}
For $d=2$, we have
\begin{equation}
2DM-{\lambda \chi M^{2}\over 4\pi}=D\oint \rho\,{\bf r}\cdot d{\bf
S}.\label{ks18}
\end{equation}
In usual situations, the term in the r.h.s. is positive (this is at
least the case for an axisymmetric distribution of particles in a
disk where it is equal to $2D\rho_{b}\pi
R^{2}$). Since $\rho\ge 0$, a sufficient condition is that ${\bf r}\cdot
d{\bf S}\ge 0$ on each point of the boundary. This criterion is
independent on the distribution itself and only depends on the domain
shape. When $\oint \rho\,{\bf r}\cdot d{\bf S}\ge 0$ \footnote{There are probably situations where this inequality is violated, in which case we expect that the value of the critical mass will differ from Eq. (\ref{ks19}). In such situations, boundary effects should play a prominent role.}, the above
relation implies that a necessary condition for the existence of
steady solutions is that
\begin{equation}
M\le M_{c}={8\pi D\over \chi\lambda}.
\label{ks19}
\end{equation}
If we introduce dimensionless parameters (or take $D=\chi=\lambda=1$), the
critical mass is simply $M_{c}=8\pi$. In terms of the critical mass
(\ref{ks19}), we can rewrite the Virial theorem (\ref{ks16}) as
\begin{equation}
{1\over 2}{dI\over dt}={\lambda \chi M\over 4\pi}(M_{c}-M)-D\oint \rho\,{\bf r}\cdot d{\bf
S} \qquad (d= 2).\label{ks20}
\end{equation}
For $M>M_c$, we have $\dot I\le \epsilon<0$ so that the moment of
inertia goes to zero in a finite time. This implies that the system
collapses to a Dirac peak at ${\bf r}={\bf 0}$ (the center of mass) in
a finite time. In an unbounded domain, assuming that $\rho$ decreases
more rapidly than $r^{-2}$ for $r\rightarrow +\infty$, the foregoing
relation reduces to
\begin{equation}
{1\over 2}{dI\over dt}={\lambda \chi M\over 4\pi}(M_{c}-M) \qquad (d= 2).\label{ks21}
\end{equation}
This relation shows that, in an unbounded domain, there can be
stationary solutions ($\dot I=0$) only when $M=M_c$. In that case, the
moment of inertia is conserved. For $M<M_c$, the moment of inertia
diverges with time $I(t)\rightarrow +\infty$ so that the system
evaporates. For $M>M_c$ the moment of inertia goes to zero
$I(t)\rightarrow 0$ in a finite time so that the system forms a Dirac
peak at ${\bf r}={\bf 0}$ in a finite time. Since the case $M=M_{c}$
lies at the frontiere between these two regimes (evaporation or
collapse) we expect that, when $M=M_{c}$, the system evolves either
towards a Dirac peak (collapse) or a completely spread profile
(evaporation).

The existence of a critical mass for chemotaxis in $d=2$ has been
found by various authors
\cite{cp,childress,nagai,herrero,crrs,biler1,dolbeault,biler2}, using
different arguments.

\section{The Smoluchowski-Poisson system}
\label{sec_sp}

In the mean field approximation and in a strong friction limit, the
dynamics of a one-component self-gravitating Brownian gas is
described by the Smoluchowski-Poisson (SP) system
\begin{equation}
{\partial\rho\over\partial t}=\nabla\cdot \biggl \lbrack
{1\over\xi}\left ({k_{B}T\over m}\nabla \rho+\rho\nabla\Phi\right )\biggr\rbrack, \label{sp1}
\end{equation}
\begin{equation}
\Delta\Phi=S_{d}G\rho. \label{sp2}
\end{equation}
The
Smoluchowski-Poisson system has been studied by Chavanis \& Sire
\cite{crs,sc,post,tcoll,banach,crrs,sopik,virial1,virial2} in
different dimensions of space (see also extensions in \cite{lang,logo}
for more general equations of state $p=p(\rho)$). Since the temperature
is fixed, the relevant statistical ensemble is the canonical
ensemble. The boundary conditions are: (i) the current ${\bf
J}={k_{B}T\over m}\nabla\rho+\rho\nabla
\Phi$ is parallel to the boundary in a finite domain and vanishes at
infinity in an unbounded domain (so that the total mass is
conserved). (ii) The gravitational potential is given by $\Phi({\bf
r},t)=-{G\over (d-2)}\int\rho({\bf r}',t)|{\bf r}-{\bf
r}'|^{-(d-2)}d{\bf r}'$ (in $d\neq 2$) or $\Phi({\bf
r},t)=-G\int\rho({\bf r}',t)\ln|{\bf r}-{\bf r}'|d{\bf r}'$ (in
$d=2$), using the usual Gauge condition.

It is clear at first sight that the Smoluchowski-Poisson system
(\ref{sp1})-(\ref{sp2}) is isomorphic to the simplified Keller-Segel
model (\ref{ks1})-(\ref{ks2}) provided that we make the
correspondances
\begin{equation}
D=\frac{k_{B}T}{\xi m}, \quad \chi=\frac{1}{\xi}, \quad c=-\Phi, \quad \lambda=S_{d}G.
\label{sp3}
\end{equation}
Furthermore, the structure of the solutions depends on a single
dimensionless parameter \cite{sc} which can be written:
\begin{equation}
\eta=\frac{\beta GM m}{R^{d-2}}=\frac{\lambda\chi M}{S_{d}R^{d-2}}.
\label{sp3new}
\end{equation}
We note, in particular, that the concentration of the chemical $c({\bf
r},t)$ in the chemotactic problem is the counterpart of the
gravitational potential $\Phi({\bf r},t)$ for self-gravitating
Brownian particles (with the opposite sign).  Due to this analogy, the
results derived for the SP system can be applied to the KS model and
vice versa.  However, due to the different notations (and also because
chemotaxis and gravity are studied by different communities) this
connection is not always made. It can be therefore of interest to put
the two models in parallel, as we do here.

The Virial theorem for the SP system has been derived in \cite{virial1} (see also Appendix \ref{sec_vtsss}). It can be written
\begin{equation}
{1\over 2}\xi{dI\over dt}=dNk_{B}T+(d-2)W-dPV, \qquad (d\neq 2)\label{sp4}
\end{equation}
where $W=\frac{1}{2}\int \rho\Phi d{\bf r}$ is the potential
energy and we have defined
\begin{equation}
P\equiv {1\over dV}\oint p\ {\bf r}\cdot d{\bf S},\label{sp5}
\end{equation}
where $p({\bf r})=\rho({\bf r}) k_{B}T/m$ is the local pressure of
an isothermal gas. In the case where $p=p_{b}$ is constant on the
boundary, we have $P=p_{b}$. In dimension $d=2$, the Virial theorem
takes the simple form \cite{virial1}:
\begin{equation}
{1\over 2}\xi{dI\over dt}=2N k_B T-{GM^{2}\over 2}-2PV, \qquad (d=2).\label{sp6}
\end{equation}
At equilibrium ($\dot I=0$), we obtain
\begin{equation}
2N k_B T-{GM^{2}\over 2}=2PV,\label{sp7}
\end{equation}
Since $P\ge 0$ in usual circumstances (this is at least the case for
an axisymmetric system in a disk where $P=\rho_{b}k_{B}T/m$), the above relation implies that a necessary condition for the
existence of steady solutions is that
\begin{equation}
T\ge  T_{c}={GMm\over 4k_{B}}.\label{sp8}
\end{equation}
If we introduce dimensionless parameters (or take $G=M=m=k_{B}=\xi=1$), the
critical temperature is simply $T_{c}=1/4$. In terms of  the critical temperature (\ref{sp8}), the Virial theorem
(\ref{sp6}) can be rewritten
\begin{equation}
{1\over 2}\xi{dI\over dt}=2N k_B (T-T_{c})-2PV.\label{sp9}
\end{equation}
For $T< T_{c}$, the moment of inertia goes to zero in a finite time
so that the system collapses to a Dirac peak at ${\bf r}={\bf 0}$ in a
finite time. At equilibrium ($\dot I=0$), we obtain
\begin{equation}
PV=Nk_B(T-T_{c}).\label{sp10}
\end{equation}
This is the equation of state of the two-dimensional self-gravitating
gas at statistical equilibrium in the thermodynamic limit
$N\rightarrow +\infty$ with $\eta= GMm/k_{B}T$ fixed (in that limit,
the mean field approximation is exact \cite{hb}). This equation of
state (and its extension to finite $N$ systems) has been obtained by
various authors using different methods
\cite{salsberg,klb,paddy,virial1,virial2,new}.

In an unbounded domain, the Virial theorem (\ref{sp9}) reduces to
\begin{equation}
{1\over 2}\xi{dI\over dt}=2N k_B (T-T_{c}).\label{sp11}
\end{equation}
This relation shows that, in an unbounded domain, there can be
stationary solutions ($\dot I=0$) only at the critical temperature
$T=T_c$. In that case, the moment of inertia is conserved.  For
$T>T_c$, the moment of inertia diverges $I(t)\rightarrow +\infty$ so
that the system evaporates. For $T<T_c$ the moment of inertia goes to
zero $I(t)\rightarrow 0$ in a finite time so that the system forms a
Dirac peak at ${\bf r}={\bf 0}$ in a finite time. Defining the
mean-squared radius of the cluster through the relation $\langle
r^{2}\rangle={I/M}$, we can rewrite Eq. (\ref{sp11}) after integration
in the form $\langle r^{2}\rangle=4D(T)t+\langle r^{2}\rangle_{0}$
where
\begin{eqnarray}
D(T)={k_{B}T\over \xi m}(1-T_{c}/T), \label{sp15}
\end{eqnarray}
is an effective diffusion coefficient. For $T\gg T_{c}$ when
gravitational effects become negligible, the Smoluchowski equation
(\ref{sp1}) reduces to a pure diffusion equation and the diffusion
coefficient is given by the Einstein formula $D(+\infty)=k_{B}T/\xi
m$. However, Eq.~(\ref{sp15}) shows that the diffusion is less and
less effective as temperature decreases and gravitational effects
come into play. In particular, the effective diffusion coefficient
becomes negative for $T<T_c$ indicating finite time collapse.

The existence of a critical temperature for self-gravitating systems
in $d=2$ has been found by various authors, in different contexts,
using different arguments \cite{ostriker,stodolkiewicz,ap,at,sc,new}. A
similar critical temperature (whose value is negative) appears in the
statistical mechanics of  point vortices in two-dimensional
hydrodynamics \cite{lp,caglioti,Chouches}.

\section{The equilibrium density profile in $d=2$}
\label{sec_ep}

\subsection{Chemotaxis of bacterial populations}
\label{sec_chemo}

The stationary solutions of the KS model (\ref{ks1})-(\ref{ks2}) are such that
\begin{equation}
\rho=Ae^{\frac{\chi}{D}c}.\label{chemo1}
\end{equation}
This is similar to the Boltzmann distribution in statistical mechanics
provided that we interprete $T_{eff}=D/\chi$ as an effective
temperature and $-c({\bf r})$ as a potential. This distribution can be
obtained by extremizing the functional
\begin{equation}
F[\rho]=-{1\over 2}\int \rho c \, d{\bf r}+{D\over\chi}\int \rho\ln\rho d{\bf r},\label{chemo2}
\end{equation}
at fixed mass. This functional is the Lyapunov functional of the KS
model. It satisfies $\dot F\le 0$ and $\dot F=0$ if, and only if, the
density is given by Eq. (\ref{chemo1}). It can also be interpreted as
an effective free energy $F=E-T_{eff}S$ associated with the Boltzmann entropy
\cite{degrad}. A steady state of the KS model is an extremum of $F$ at
fixed mass. Furthermore, it is linearly dynamically stable if and only
if it is a (local) {\it minimum} of this functional \cite{gen}. The
equilibrium state is obtained by substituting Eq. (\ref{chemo1}) into
the Poisson equation (\ref{ks2}) yielding
\begin{equation}
\Delta c=-\lambda A e^{\frac{\chi}{D}c}.\label{chemo3}
\end{equation}
This is similar to the Boltzmann-Poisson equation appearing in
astrophysics (in the statistical mechanics of stellar systems
\cite{paddy,ijmpb} and in isothermal models of stars \cite{chandra}) and in vortex dynamics
(in the statistical mechanics of 2D point vortices \cite{Chouches}). If
we restrict ourselves to axisymmetric solutions, this equation can be
solved analytically in $d=2$. The 2D axisymmetric Boltzmann-Poisson
equation (\ref{chemo3}) is solved in Appendix \ref{sec_bp}. Here we
use a slightly different method working directly on the mass profile.

Considering axisymmetric density profiles, introducing the
accumulated mass $M(r,t)=\int_{0}^{r}\rho(r',t)2\pi r' dr'$, using
the relation $\frac{\partial c}{\partial r}=-\frac{\lambda
M(r,t)}{2\pi r}$ (equivalent to the Gauss theorem) and introducing
the variable $u=r^{2}$, it is found that the KS model
(\ref{ks1})-(\ref{ks2}) is equivalent to the single partial
differential equation
\begin{equation}
\frac{\partial M}{\partial t}=4Du\frac{\partial^{2}M}{\partial u^{2}}+\frac{\lambda\chi}{\pi}M\frac{\partial M}{\partial u}.
\label{chemo4}
\end{equation}
The stationary profiles satisfy
\begin{equation}
uM''+\frac{\lambda\chi}{4\pi D}MM'=0.
\label{chemo5}
\end{equation}
Using $uM''=(uM')'-M'$, $2MM'=(M^{2})'$ and $M(0)=0$, we obtain after
integration the first order differential equation
\begin{equation}
uM'=M\left (1-\frac{\lambda\chi}{8\pi D}M\right ).
\label{chemo6}
\end{equation}
Since $M'\ge 0$, a necessary condition for the existence of steady states is that
\begin{equation}
M\le M_{c}={8\pi D\over \chi\lambda}.
\label{chemo7}
\end{equation}
Solving Eq. (\ref{chemo6}) when this condition is fulfilled, we find
that the equilibrium mass profile is given by
\begin{equation}
M(r)=\frac{Kr^{2}}{1+\frac{Kr^{2}}{M_{c}}},
\label{chemo8}
\end{equation}
where $K$ is a constant of integration determined by the boundary
conditions.

\begin{figure}
\vskip1cm
\centerline{
\psfig{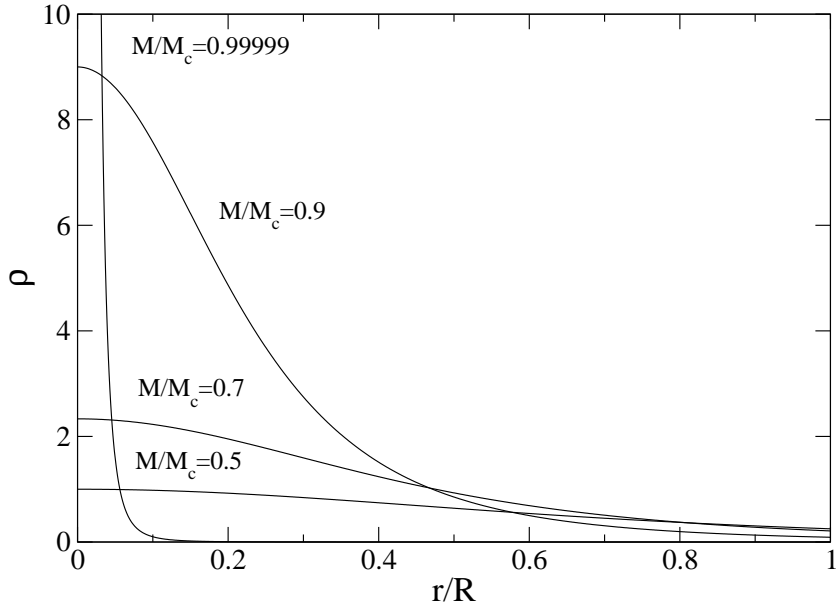} }
\caption{Density profile (normalized by $M_{c}/\pi R^{2}$) as a function of $r/R$ for different values of the total mass $M$. For $M\rightarrow M_{c}$, the density profile tends to a Dirac peak.}
\label{profilsdensite}
\end{figure}

\begin{figure}
\vskip1cm
\centerline{
\psfig{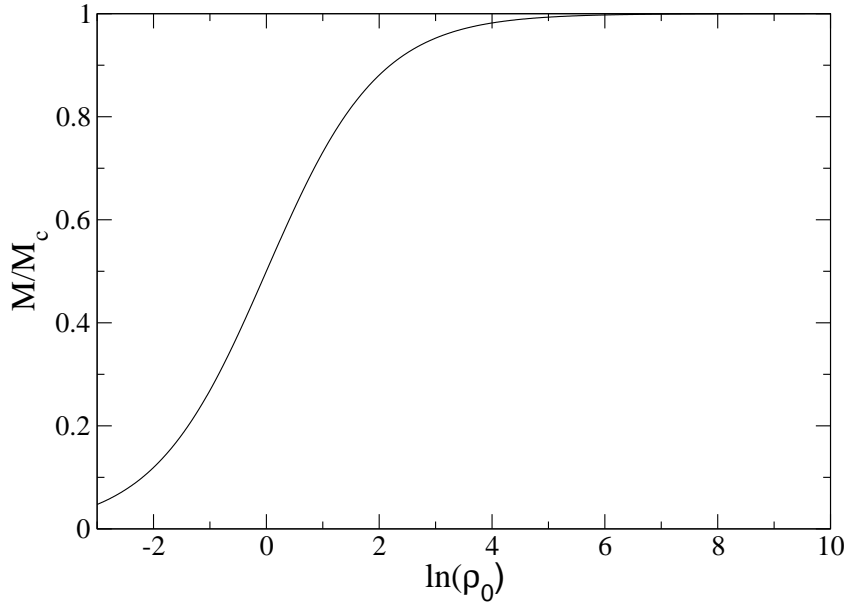}} \caption{Relation between the mass $M/M_{c}$ and the central density $\rho_{0}$ (normalized by $M_{c}/\pi R^{2}$) in the chemotactic problem. There exists a unique solution for each value of $M\le M_{c}$. Since there is no turning point in the series of equilibria $M(\rho_{0})$, these steady solutions are stable (minima of free energy $F$ at fixed mass $M$) \cite{sc}.}
\label{massdensite}
\end{figure}

(a) In a disk of radius $R$, the constant $K$ is determined by the condition $M(R)=M$. This yields
\begin{equation}
M(r)=\frac{M}{1-\frac{M}{M_{c}}}\frac{(\frac{r}{R})^{2}}{1+\frac{M}{M_{c}-M}(\frac{r}{R})^{2}}.
\label{chemo9}
\end{equation}
The corresponding density profile is
\begin{equation}
\rho(r)=\frac{1}{\pi R^{2}}\frac{M}{1-\frac{M}{M_{c}}}\frac{1}{\left \lbrack 1+\frac{M}{M_{c}-M}(\frac{r}{R})^{2}\right \rbrack^{2}}.
\label{chemo10}
\end{equation}
Typical density profiles are plotted in Fig. \ref{profilsdensite} for
different values of the ratio $M/M_{c}$.  The relation between the
mass and the central density is
\begin{equation}
\rho_{0}=\frac{M_{c}}{\pi R^{2}}\frac{M/M_{c}}{1-\frac{M}{M_{c}}}.
\label{chemo11}
\end{equation}
It is plotted in Fig. \ref{massdensite}. The central density increases as $M\rightarrow M_{c}$. For $M=M_{c}$,
the density profile is a Dirac peak: $\rho({\bf r})=M_{c}\delta({\bf
r})$.

(b) In an unbounded domain, we see from Eq. (\ref{chemo8}) that the
condition $M(r)\rightarrow M$ for $r\rightarrow +\infty$ requires that
$M=M_{c}$. Therefore, there exists steady state solutions only for
$M=M_{c}$. They form an infinite family of solutions parameterized by the
constant $K$ or equivalently by the central density
$\rho_{0}=K/\pi$. Their density profile is
\begin{equation}
\rho(r)=\frac{\rho_{0}}{\left (1+\frac{\pi\rho_{0}r^{2}}{M_{c}}\right )^{2}}.
\label{chemo12}
\end{equation}
It is suggested in Sec. \ref{sec_ks} that, among all the
distributions of this family, the only stable stationary solution of
the KS model with $M=M_{c}$ is the Dirac peak corresponding to
$\rho_{0}\rightarrow +\infty$. The dynamical formation of this Dirac
peak is shown analytically in \cite{virial1}. Another possible
evolution is an evaporation.

\subsection{Self-gravitating Brownian particles}
\label{sec_brown}

We give, for comparison, the equivalent relations for the
gravitational problem. The stationary solutions of the Smoluchowski
equation (\ref{sp1}) are described by the Boltzmann distribution
\begin{equation}
\rho=Ae^{-\beta m\Phi}.\label{brown1}
\end{equation}
This is equivalent to the condition of hydrostatic equilibrium $\nabla
p+\rho\nabla\Phi={\bf 0}$ for an isothermal equation of state $p({\bf
r})=\rho({\bf r})k_{B}T/m$ \cite{virial1}. 
The Lyapunov functional of
the SP system is the Boltzmann free energy
\begin{equation}
F[\rho]={1\over 2}\int \rho\Phi \, d{\bf r}+k_{B}T\int {\rho\over m}\ln {\rho\over m}\, d{\bf r}. \label{brown2}
\end{equation}
The steady states  are obtained by solving
the Boltzmann-Poisson equation
\begin{equation}
\Delta \Phi=S_{d}G A e^{-\beta m\Phi}.\label{brown3}
\end{equation}
They are linearly dynamically stable if, and only if,
they are {\it minima} of free energy at fixed mass. This is consistent
with a condition of thermodynamical stability in the canonical
ensemble. In $d=2$, the SP system is equivalent to a single partial
differential equation for the mass profile
\begin{equation}
\xi\frac{\partial M}{\partial t}=4Tu\frac{\partial^{2}M}{\partial u^{2}}+2GM \frac{\partial M}{\partial u}.
\label{brown4}
\end{equation}
(a) In a bounded domain, there exists steady states if, and only, if
\begin{equation}
T\ge  T_{c}={GMm\over 4k_{B}}.\label{brown5}
\end{equation}
The mass profile is given by
\begin{equation}
M(r)=\frac{M}{1-\frac{T_{c}}{T}}\frac{(\frac{r}{R})^{2}}{1+\frac{T_{c}}{T-T_{c}}(\frac{r}{R})^{2}},
\label{brown6}
\end{equation}
and the corresponding density profile by
\begin{equation}
\rho(r)=\frac{1}{\pi R^{2}}\frac{M}{1-\frac{T_c}{T}}\frac{1}{\left\lbrack 1+\frac{T_{c}}{T-T_{c}}(\frac{r}{R})^{2}\right\rbrack^{2}}.
\label{brown7}
\end{equation}
The relation between the temperature  and the central density is
\begin{equation}
\rho_{0}=\frac{1}{\pi R^{2}}\frac{M}{1-\frac{T_c}{T}}.
\label{brown8}
\end{equation}
At  $T=T_{c}$, the density is a Dirac peak $\rho({\bf r})=M\delta({\bf r})$.

(b) In an unbounded domain, there exists steady solutions only for  $T=T_{c}$.  They are parameterized by  the central density $\rho_{0}=K/\pi$. The mass and density profiles are
\begin{equation}
M(r)=\frac{Kr^{2}}{1+\frac{K}{M}r^{2}},\qquad \rho(r)=\frac{\rho_{0}}{\left (1+\frac{\pi\rho_{0}}{M}r^{2}\right )^{2}}.
\label{brown10}
\end{equation}
These solutions have the same value of the free energy
$F=-Nk_{B}T_{c}[1+\ln(\pi/N)]$ independent on $\rho_{0}$. On the other
hand, their moment of inertia is infinite (except the solution with
$\rho_{0}=+\infty$ for which $I=0$). Since $I$ is conserved at
$T=T_{c}$, they cannot be reached from a generic initial condition
with $0<I<+\infty$. It is suggested in Sec. \ref{sec_sp} that the only
stable stationary solution of the SP system with $T=T_{c}$ is the
Dirac peak corresponding to $\rho_{0}\rightarrow +\infty$. The
dynamical formation of this Dirac peak is shown analytically in
\cite{virial1}. The system ejects a tiny amount of mass at large
distances in order to accomodate for the conservation of $I$. Another
possible evolution at $T=T_{c}$ is an evaporation.

In the above formulae, we can pass from biological notations to  gravitational notations by using the relation
\begin{equation}
\frac{T}{T_{c}}=\frac{M_{c}}{M},
\label{brown11}
\end{equation}
resulting from the correspondances (\ref{sp3}). Therefore, the mass
$M$ in chemotaxis plays the role of the inverse temperature
$\beta=1/k_B T$ in gravity. We note also that, in 2D gravity, the
condition (\ref{brown5}) for the existence of a steady solution above
a critical temperature $T_{c}$ (for a given mass $M$), can be converted
into a condition on the mass (for a given temperature $T$)
\begin{equation}
M\le  M_{c}={4k_{B}T\over Gm},\label{brown12}
\end{equation}
making the link with the biological problem even closer.

\section{Analogy with the Burgers equation in $d=1$}
\label{sec_ab}

\subsection{The case of an infinite domain}
\label{sec_one}

In $d=1$, it is shown in \cite{sc} that the Smoluchowski-Poisson system is equivalent to a single differential equation
\begin{equation}
\xi\frac{\partial M}{\partial t}=\frac{k_{B}T}{m}\frac{\partial^{2}M}{\partial x^{2}}+GM\frac{\partial M}{\partial x},\label{one1}
\end{equation}
for the integrated density $M(x,t)=2\int_{0}^{x}\rho(x',t)dx'$. The
density and the gravitational potential are related to the mass
profile by $\partial_{x}M=2\rho(x,t)$ and
$\partial_{x}\Phi=GM(x,t)$. The stationary states of Eq.
(\ref{one1}) can be obtained analytically (see Sec. \ref{sec_onefin}). In an
infinite domain, they are given by
\begin{equation}
M(x)=M\tanh(x/H), \qquad \rho(x)=\frac{\rho_{0}}{\cosh^{2}(x/H)},\label{one2}
\end{equation}
where the  length scale and the central density are determined as a function of the temperature by
\begin{equation}
H=\frac{2k_{B}T}{GMm}, \qquad \rho_{0}=\frac{M}{2H}=\frac{GM^{2}m}{4k_{B}T}.\label{one3}
\end{equation} 
There exists a unique steady state solution for each value of the
temperature. Introducing rescaled parameters, or taking equivalently
$\xi=k_{B}=m=G=M=1$, the dynamical equation (\ref{one1}) takes the
form
\begin{equation}
\frac{\partial M}{\partial t}=T\frac{\partial^{2}M}{\partial x^{2}}+M\frac{\partial M}{\partial x}.\label{one4}
\end{equation}
Setting $v(x,t)=-M(x,t)$ and $\nu=T$, it becomes equivalent to the one dimensional Burgers equation
\begin{equation}
\frac{\partial v}{\partial t}+v\frac{\partial v}{\partial x}=\nu\frac{\partial^{2}v}{\partial x^{2}},\label{one5}
\end{equation}
which has been studied in connection with hydrodynamical turbulence
\cite{burgers} and cosmology \cite{frisch}.  In this analogy, the mass
plays the role of the velocity ($v=-M$), the density the role of the
velocity increment ($v'=-2\rho$) and the temperature the role of the
viscosity ($\nu=T$). The steady solution of Eq. (\ref{one5}) is
\begin{equation}
v(x)=-\tanh(x/2\nu).\label{one6}
\end{equation}
In 1D turbulence, this solution describes a single shock
\cite{burgers} while in the context of self-gravitating Brownian
particles it describes the equilibrium mass profile (\ref{one2}) of a
one dimensional isothermal gas \cite{virial1}.

The collapse dynamics of the Smoluchowski-Poisson system at $T=0$ has
been studied in \cite{sc,post} where an analytical solution describing
the formation of a Dirac peak has been obtained.  The collapse is
self-similar and a Dirac peak is formed in the post-collapse regime
\cite{sc,post}.  This Dirac peak is precisely the limiting form of the
stationary solution (\ref{one2})-(\ref{one3}) at $T=0$. We note that
the solution obtained in \cite{sc,post} also describes the formation
of a singular shock in inviscid 1D turbulence ($\nu=0$) since the
equation for the mass profile of self-gravitating Brownian particles
in $d=1$ is isomorphic to the Burgers equation.  For $\nu\neq 0$, the
Burgers equation (\ref{one5}) has a well-known explicit solution. As
observed by Hopf
\cite{hopf} and Cole \cite{cole}, the change of variables
$v=-\partial_{x}\psi$ and $\psi=2\nu\ln\theta$ transforms the
nonlinear Burgers equation into the heat equation thereby leading to
the explicit solution \cite{frisch}:
\begin{equation}
\psi(x,t)=2\nu\ln\left\lbrace \frac{1}{\sqrt{4\pi \nu t}}\int_{-\infty}^{+\infty}{\rm exp}\left\lbrack \frac{1}{2\nu}\left (\psi_{0}(q)-\frac{(x-q)^{2}}{2t}\right )\right\rbrack dq\right\rbrace, \label{one7}
\end{equation}
where $\psi_{0}(x)=\psi(x,0)$. Returning to the notations of the
initial problem, we note that $\psi$ represents the gravitational
potential $\Phi$. This provides the general solution of the SP system
in $d=1$ dimension in an unbounded domain for $T\neq 0$. Acedo \cite{acedo} has
constructed an explicit analytical solution of the 1D
Smoluchowski-Poisson system from the general formula (\ref{one7}). If
we take the limit $T\rightarrow 0$ and use steepest descent technics,
we obtain from Eq. (\ref{one7}):
\begin{equation}
\psi(x,t)={\rm sup}_{q}\left \lbrack \psi_{0}(q)-\frac{(x-q)^{2}}{2t}\right\rbrack, \qquad (T\rightarrow 0). \label{one8}
\end{equation}
A general method to solve this equation is described in
\cite{frisch} in the cosmological context. Interestingly,
these results can also have applications for the 1D Smoluchowski-Poisson
system.

\subsection{The case of a finite domain}
\label{sec_onefin}

\begin{figure}
\vskip1cm
\centerline{
\psfig{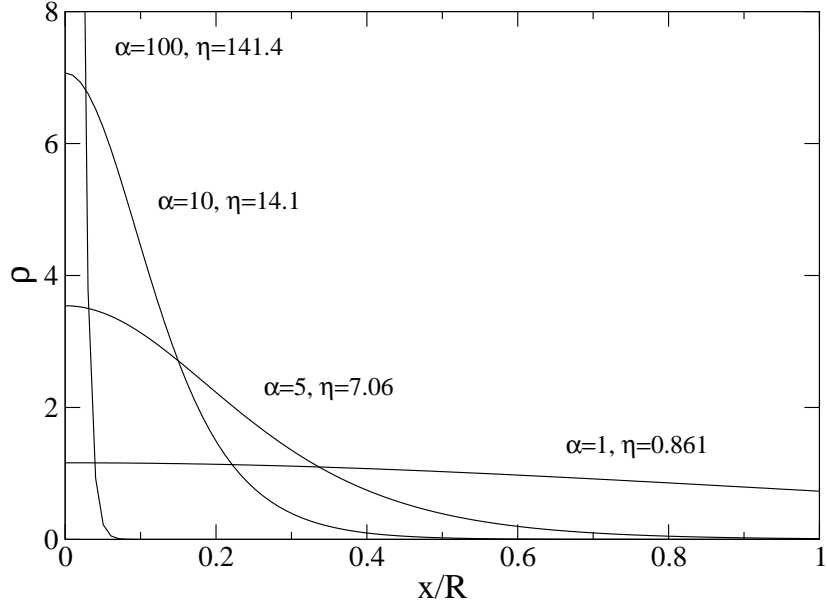} }
\caption{Density profile (normalized by $M/2R$) as a function of $x/R$ for different values of the temperature $\eta=\beta GMmR$. For $T\rightarrow 0$, the density profile tends to a Dirac peak.}
\label{rhoxd1}
\end{figure}

According to Eq. (\ref{one1}), the stationary solution of the SP system in $d=1$ satisfies the
differential equation
\begin{equation}
M''+\frac{Gm}{k_B T}M M'=0. \label{exunm1}
\end{equation}
Integrating this equation once and using $M(0)=0$ and
$M'(0)=2\rho_0$ to determine the constant of integration, we find
that
\begin{equation}
M'+\frac{Gm}{2 k_B T}M^2=2\rho_0. \label{exunm2}
\end{equation}
This can be rewritten
\begin{equation}
\frac{dM}{a^2-M^2}=\frac{Gm}{2 k_B T}dx, \qquad a=\left (\frac{4k_B
T\rho_0}{Gm}\right )^{1/2}. \label{exunm3}
\end{equation}
This is easily integrated in
\begin{equation}
M(x)=a\tanh\left (\frac{Gma}{2k_B T}x\right ), \qquad \rho(x)=\frac{\rho_{0}}{\cosh^{2}\left (\frac{Gma}{2k_{B}T}x\right )}.\label{exunm4}
\end{equation}
In an infinite domain, using the fact that $M(x)\rightarrow M$ for
$x\rightarrow +\infty$, we find that $a=M$ and we obtain the relations
(\ref{one2})-(\ref{one3}). In a bounded domain, using the fact that
$M(R)=M$ we find that $a$ is solution of
\begin{equation}
M=a\tanh\left (\frac{G m a R}{2k_B T}\right ).\label{exunm5}
\end{equation}
This equation implicitly determines the central density $\rho_{0}$ in
terms of the temperature $T$. These results can also be obtained by
solving the 1D Boltzmann-Poisson equation (see Appendix
\ref{sec_bpun}) which is equivalent to Eq. (\ref{exunm1}) \cite{sc}. 
The equilibrium density profile for different values of the
temperature is represented in Fig. \ref{rhoxd1} and the relation
between the temperature and the central density is represented in
Fig. \ref{etarhod1}. We have used the notations defined in Appendix
\ref{sec_bpun}.

\begin{figure}
\vskip1cm
\centerline{
\psfig{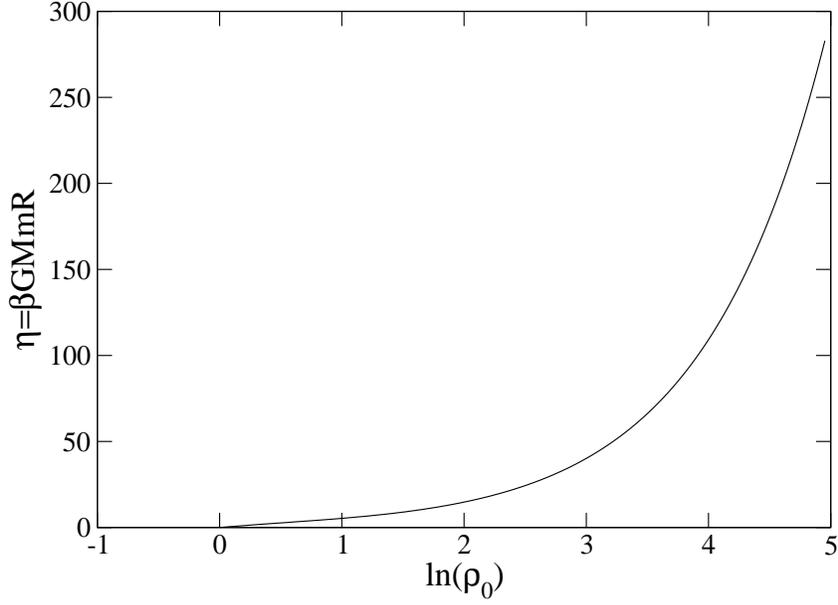}} \caption{Relation between the normalized inverse temperature $\eta=\beta GMmR$ and the central density $\rho_{0}$ (normalized by $M/2 R$). There exists a unique solution for each value of the temperature. Since there is no turning point in the series of equilibria $\eta(\rho_{0})$, these steady solutions are stable (minima of free energy $F$ at fixed mass $M$) \cite{sc}. }
\label{etarhod1}
\end{figure}

Introducing rescaled variables or taking equivalently
$\xi=k_B=m=G=M=R=1$, the dynamical equation for the mass profile in
$d=1$ is given by Eq. (\ref{one4}) with the boundary conditions
$M(0,t)=0$, $M(1,t)=1$ and $M(x,0)=M_0(x)$. With the change of
variables $v(x,t)=-M(x,t)$, $v=-\partial_x\psi$ and $\psi=2T\ln\theta$
of Sec. \ref{sec_one}, we find that the function $\theta(x,t)$
satisfies the diffusion equation
\begin{equation}
\frac{\partial\theta}{\partial t}=T\frac{\partial^2\theta}{\partial
x^2},\label{dynun1}
\end{equation}
with the boundary conditions
\begin{equation}
\theta'(0,t)=0, \quad \theta'(1,t)=\frac{1}{2T}\theta(1,t), \quad
\theta_0(x)=e^{\frac{1}{2T}\int_{0}^{x}M_0(y)dy}.\label{dynun2}
\end{equation}
Using the method of separation of variables, we find that the
general solution of these equations is
\begin{equation}
\theta(x,t)=a_0 e^{T\lambda_0^2 t}\cosh(\lambda_0
x)+\sum_{n=1}^{+\infty}a_n e^{-T\lambda_n^2 t}\cos(\lambda_n x),
\label{dynun3}
\end{equation}
where $\lambda_0$ is the solution of
$\tanh(\lambda_0)=1/(2T\lambda_0)$ and $\lambda_n$ are the solutions
 of $\tan(\lambda_n)=-1/(2T\lambda_n)$. The coefficients are
 determined from the initial condition
\begin{equation}
\theta_0(x)=a_0 \cosh(\lambda_0 x)+\sum_{n=1}^{+\infty}a_n
\cos(\lambda_n x). \label{dynun4}
\end{equation}
Using the relations satisfied by $\lambda_0$ and $\lambda_n$, the
following identities can be obtained by standard calculations
\begin{equation}
\int_0^1 \cos(\lambda_n x)\cos(\lambda_m x)
dx=\frac{1+4T^2\lambda_n^2-2T}{2(1+4T^2\lambda_n^2)}\delta_{nm},
\label{dynun5}
\end{equation}
\begin{equation}
\int_0^1 \cosh(\lambda_0 x)\cos(\lambda_n x) dx=0, \label{dynun6}
\end{equation}
\begin{equation}
\int_0^1 \cosh^2(\lambda_0 x)
dx=\frac{4T^2\lambda_0^2-1+2T}{2(4T^2\lambda_0^2-1)}. \label{dynun7}
\end{equation}
Then, we find that the coefficients $a_{n}$ are given in terms of the
initial condition by
\begin{equation}
a_0=\frac{2(4T^2\lambda_0^2-1)}{4T^2\lambda_0^2-1+2T}\int_0^1
\theta_0(x) \cosh(\lambda_0 x) dx, \label{dynun8}
\end{equation}
\begin{equation}
a_n=\frac{2(1+4T^2\lambda_n^2)}{1+4T^2\lambda_n^2-2T}\int_0^1
\theta_0(x) \cos(\lambda_n x) dx. \label{dynun9}
\end{equation}
The foregoing equations provide the general solution of the SP system
in $d=1$ in a bounded domain. For $t\rightarrow +\infty$, we find from
Eq. (\ref{dynun3}) that $M(x,t)\rightarrow M(x)=2T\lambda_0
\tanh(\lambda_0 x)$ with $1=2T\lambda_0\tanh(\lambda_0)$ which returns
the stationary solution (\ref{exunm4})-(\ref{exunm5}).

\subsection{Generalization to other dimensions}
\label{sec_onegen}

Note, finally, that the Smoluchowski-Poisson system in $d$ dimensions is equivalent to the single differential equation \cite{sc}:
\begin{equation}
\xi\frac{\partial M}{\partial t}=\frac{k_{B}T}{m}\left (\frac{\partial^{2}M}{\partial r^{2}}-\frac{d-1}{r}\frac{\partial M}{\partial r}\right )+\frac{G}{r^{d-1}}M\frac{\partial M}{\partial r},\label{one9}
\end{equation}
for the mass profile $M(r,t)=\int_{0}^{r}\rho(r',t)S_{d}r^{'d-1}dr'$. With the change of variables $x=r^{d}/d$, the foregoing equation becomes
\begin{equation}
\xi\frac{\partial M}{\partial t}=\frac{k_{B}T}{m}(dx)^{\frac{2(d-1)}{d}}\frac{\partial^{2}M}{\partial x^{2}}+GM\frac{\partial M}{\partial x}.\label{one10}
\end{equation}
Introducing rescaled variables, or using equivalently
$\xi=k_B=m=G=M=R=1$, and setting $v(x,t)=-M(x,t)$, this can be put in
the form of a Burgers equation:
\begin{equation}
\frac{\partial v}{\partial t}+v\frac{\partial v}{\partial x}=\nu(x)\frac{\partial^{2}v}{\partial x^{2}},\label{one11}
\end{equation}
with a position dependent viscosity
\begin{equation}
\nu(x)=T(d\ x)^{\frac{2(d-1)}{d}}.\label{one12}
\end{equation}
The viscosity is a pure power-law scaling like $1$, $x$ and
$x^{4/3}$ in $d=1,2$ and $3$ dimensions respectively. For $T=0$, we
obtain the inviscid Burgers equation in any dimension of space.

In conclusion, it is interesting to note that the spherically
symmetric Smoluchowski-Poisson system is connected to the 1D Burgers
equation which arises in many domains of physics. This increases the
interest of the self-gravitating Brownian gas model \cite{crs}. We
note however that, in our case, $v(x,t)=-M(x,t)$ is a negative
quantity for $x\ge 0$ while in 1D turbulence the velocity $v$ can take
positive and negative values.

\section{Dynamics of bacterial populations and self-gravitating Brownian particles}
\label{sec_dyn}

The Virial theorem (\ref{sp4}) can be used to obtain general results
on the dynamics of self-gravitating Brownian particles without
solving the equations of motion.  However, the explicit
resolution of the Smoluchowski-Poisson system, as considered by
Chavanis \& Sire
\cite{crs,sc,post,tcoll,banach,crrs,sopik,virial1,virial2}, gives more precise information on the evolution of the
density profile of the particles. In this section, we provide a
summary of these results in the case of  self-gravitating Brownian
particles and transpose them to the chemotactic problem by using the
notations of biology. This should reinforce the link between these
two topics that are studied by different communities.

\subsection{Self-gravitating Brownian particles}
\label{sec_qualb}

First, consider a two-dimensional self-gravitating Brownian gas
enclosed within a circular box of radius $R$. (a) Regular steady
states exist for $T>T_{c}$ and their density profile is known
analytically, see Eq. (\ref{brown7}).  They are global minima of the
free energy (\ref{brown2}) at fixed mass. Using the free energy as a
Lyapunov functional, we deduce that the SP system relaxes
towards these steady states.  The particles are confined by the box
since $P>0$ (i.e.  $\rho(R)>0$). (b) For $T=T_{c}$, the steady state
is a Dirac peak $\rho({\bf r})=M\delta({\bf r})$ containing the whole
mass so that $P=0$ (i.e.  $\rho(R)=0$). The dynamics of the SP system
at $T=T_{c}$ has been studied in Sec. IV.B. of \cite{sc}. The collapse
is self-similar with a scaling function similar to Eq. (\ref{brown7}),
and the Dirac peak containing the whole mass is formed for
$t\rightarrow +\infty$.  The central density increases exponentially
rapidly as $\rho_{0}(t)\sim e^{\sqrt{2t}}$. (c) For $T<T_{c}$, there
is no steady state and the system undergoes gravitational
collapse. The dynamics of the SP system at $T<T_{c}$ has been studied
in Sec.  IV.C. of \cite{sc}. For $T<T_{c}$ the collapse is not exactly
self-similar. In a finite time $t_{coll}$, the system develops a Dirac
peak containing a fraction $T/T_{c}$ of the total mass $M$, surrounded
by a halo whose tail decreases as $\rho\sim r^{-\alpha(t)}$ with
$\alpha(t)$ converging {\it extremely slowly} to $\alpha=2$ for
$t\rightarrow t_{coll}$, like
$2-\alpha(t)=\sqrt{2\ln\ln\rho_{0}(t)/\ln\rho_{0}(t)}$ with
$\rho_{0}(t)\sim (t_{coll}-t)^{-1}$, so that the evolution {\it looks}
self-similar. For $t\rightarrow t_{coll}$, the density profile behaves
like $\rho({\bf r},t)\rightarrow (T/T_{c})M\delta({\bf r})+\tilde
\rho({\bf r},t)$ where the evolution of $\tilde \rho({\bf r},t)$ is
studied in \cite{sc}. From the Virial theorem Eq.~(\ref{sp9}), we note
that $\dot I\le \epsilon<0$ for $T<T_{c}$ so that the moment of
inertia tends to its minimum value $I=0$ in a finite time
$t_{end}$. This corresponds to the formation of a Dirac peak at ${\bf
r}={\bf 0}$ containing the whole mass $M$. Since this final state is
different from the structure obtained at $t=t_{coll}$, this means that
the evolution continues in the post-collapse regime $t_{coll}\le t\le
t_{end}$.  In this post-collapse regime, the Dirac peak formed at
$t=t_{coll}$ accretes the mass of the surrounding halo until all the
mass is at ${\bf r}={\bf 0}$ at $t_{end}$. This regime has not yet been
fully described.

Consider now a two-dimensional self-gravitating Brownian gas in an
infinite domain. (a) For $T>T_{c}$, the particles have a diffusive
motion (evaporation) with an effective diffusion coefficient given by
Eq. (\ref{sp15}) increasing linearly with the distance $T-T_{c}$ to
the critical temperature. This evaporation process is self-similar and
it has been analytically studied in Secs. IV.B. and IV.C.  of
\cite{virial1} for $T\gg T_{c}$ and $T\rightarrow T_{c}^{+}$
respectively. (b) For $T=T_{c}$, the effective diffusion coefficient
vanishes $D(T_{c})=0$, so that the moment of inertia is conserved:
$\dot I=0$. The dynamics of the SP system at $T=T_{c}$ has been
studied in Sec. D. of \cite{virial1}. There exists an infinite family
of steady state solutions (\ref{brown10}) parameterized by the central density
$\rho_{0}$ but they have $I=+\infty$ (for $\rho_{0}<+\infty$) and
cannot be reached by the system for generic initial conditions (with
$I<+\infty$). An analytical dynamical solution has been found in which the
system forms a Dirac peak of mass $M-\epsilon$ for $t\rightarrow
+\infty$ and ejects a tiny amount of mass $\epsilon\ll 1$ at large
distances so as to satisfy the moment of inertia constraint (if $I\neq
0$ initially).  Note that in an unbounded domain
\cite{virial1}, the central density diverges logarithmically with time,
$\rho_{0}(t)\sim \ln t$, while the divergence is exponential in a
bounded domain \cite{sc}. Another possible evolution at $T=T_{c}$ is
an evaporation. (c) For $T<T_{c}$, the effective diffusion coefficient
in Eq. (\ref{sp15}) is negative, implying finite time blow-up. In
particular, $\langle r^{2}\rangle=0$ for $t_{end}=\langle
r^{2}\rangle_{0}/4|D(T)|$ where $\langle r^{2}\rangle_{0}$ is
calculated at $t=0$ from the center of mass. This leads to the
formation of a Dirac peak $\rho({\bf r})=M\delta({\bf r})$ at ${\bf
r}={\bf 0}$ containing the whole mass. The solution obtained in
\cite{sc} in a bounded domain probably describes the collapse of the
core accurately but, in the absence of a confining box, the collapse
is accompanied by an unlimited expansion of the halo. In the
post-collapse regime, the Virial theorem indicates that all the matter
falls at ${\bf r}={\bf 0}$ in a finite time.

For comparison, let us describe the situation in $d=3$. When the
system is confined within a spherical box of radius $R$, there exists
a critical temperature $k_{B}T_{c}={GMm\over 2.52 R}$ which depends on
the box radius $R$ (contrary to the case $d=2$). (a) For $T\ge T_{c}$,
there exists equilibrium states. They have a density contrast
$\rho(0)/\rho(R)\le 32.1$.  These equilibrium states are {\it
metastable} \footnote{Since there is no global minimum of free energy
for an isothermal self-gravitating gas, these equilibrium states may
not always be reached by the system. For some peculiar initial
conditions, the system may rather collapse and form a singularity
(Dirac peak) with infinite free energy as in the case $T<T_{c}$ (see
below). This depends on a complicated notion of basin of attraction as
illustrated in \cite{crs}. However, for generic initial conditions
with $T\ge T_{c}$, the SP system usually converges towards a
``gaseous'' steady state with smooth density profile (local minimum of
free energy). }, i.e. local minima of the free energy (\ref{brown2})
at fixed mass \cite{crs,sc}.  For $T=T_{c}$, the equilibrium state is
not a Dirac peak (contrary to the case $d=2$). The density profile is
regular with a density contrast $\rho(0)/\rho(R)=32.1$. (b) For
$0<T<T_{c}$, there is no equilibrium state and the system undergoes a
self-similar collapse $\rho(r,t)=\rho_{0}(t)f(r/r_{0}(t))$ leading to
a finite time singularity at $t=t_{coll}$. The collapse time behaves
with the distance to the critical point as
$t_{coll}=t_{*}(\eta-\eta_{c})^{-1/2}$ where $\eta=\beta GMm/R$ and
$t_{*}=0.91767702...$ \cite{tcoll}. This pre-collapse can be described
analytically in $d>2$ and the invariant density profile is known
exactly; the density decreases as $r^{-2}$ at large distances
\cite{crs,sc}. The central density increases with time as
$\rho_{0}\sim (t_{coll}-t)^{-1}$ and the core radius decreases as
$r_{0}(t)\sim (t_{coll}-t)^{1/2}$. Therefore, this pre-collapse regime
does not create a core since the mass $M_{0}(t)\sim \rho_{0}r_{0}^{d}$
at $r=0$ goes to zero as $M_{0}(t)\sim (t_{coll}-t)^{(d-2)/2}$ at
$t=t_{coll}$. However, a Dirac peak is formed in the post-collapse
regime for $t>t_{coll}$
\cite{post}. The mass of the Dirac peak increases as $M_{0}(t)\sim
(t-t_{coll})^{(d-2)/2}$ for $t\rightarrow t_{coll}^{+}$ and the residual
density obeys a backward dynamical scaling
$\rho(r,t)=\rho_{0}(t)g(r/r_{0}(t))$ with $\rho_{0}\sim
(t-t_{coll})^{-1}$ and $r_{0}\sim (t-t_{coll})^{1/2}$. The Dirac peak
accrets all the mass in an infinite time $t\rightarrow +\infty$. (c)
For $T=0$, the system undergoes a self-similar collapse and develops a
finite time singularity at $t=t_{coll}$ which can be studied
analytically in any dimension. The invariant density profile is
solution of an implicit equation and the density decreases as
$r^{-2d/(d+2)}$ at large distances
\cite{crs,sc}. The central density increases with time as
$\rho_{0}\sim (t_{coll}-t)^{-1}$ and the core radius decreases as
$r_{0}(t)\sim (t_{coll}-t)^{(d+2)/2d}$. Therefore, this pre-collapse
regime does not create a core since the mass $M_{0}(t)\sim
(t_{coll}-t)^{d/2}$ at $r=0$ goes to zero at $t_{coll}$. However, a
Dirac peak is formed in the post-collapse regime for $t>t_{coll}$
\cite{sc,post}. The mass of the Dirac peak increases as $M_{0}(t)\sim
(t-t_{coll})^{d/2}$ for $t\rightarrow t_{coll}^{+}$ and the residual
density obeys a backward dynamical scaling
$\rho(r,t)=\rho_{0}(t)g(r/r_{0}(t))$ with $\rho_{0}\sim
(t-t_{coll})^{-1}$ and $r_{0}\sim (t-t_{coll})^{(d+2)/2d}$. At $T=0$,
the Dirac peak accretes all the mass in a finite time $t_{end}=1/d$
\cite{post}.

For 3D self-gravitating Brownian particles in an unbounded domain,
there is no steady state. The system can either collapse (as in a
bounded domain) or evaporate. The evaporation process has been treated
in Sec. V. of \cite{virial1} and the collapse in \cite{crs,sc}. The
choice between collapse or evaporation depends on a complicated notion
of basin of attraction that is function of the initial condition.

Finally, in $d=1$ the problem can be mapped on the Burgers equation
which has the explicit solution (\ref{one7}) in an unbounded domain
and (\ref{dynun3}) in a bounded domain.

\subsection{Chemotactic aggregation of bacterial populations}
\label{sec_isochem}

The reduced Keller-Segel model (\ref{ks1})-(\ref{ks2}) and the
Smoluchowski-Poisson system (\ref{sp1})-(\ref{sp2}) are isomorphic
so we can directly transpose the results obtained for
self-gravitating Brownian particles to the biological context. By
reformulating these results with notations appropriate to
chemotaxis, we aim to facilitate the comparison with the numerous
results obtained in chemotaxis by applied mathematicians
\cite{horstmann}. Since we are just transposing the results of the
previous section to a different  context, we shall give less details
in their description.

First, consider the two-dimensional axisymmetric Keller-Segel model
(\ref{ks1})-(\ref{ks2}) in a circular domain of radius $R$. (a)
Regular steady states exist for $M<M_{c}$ and their density profile
is known analytically, see Eq. (\ref{chemo10}).  (b) For $M=M_{c}$,
the steady state is a Dirac peak $\rho({\bf r})=M_{c}\delta({\bf
r})$ containing the whole mass. The collapse is self-similar and the
Dirac peak is formed for $t\rightarrow +\infty$. The central density
increases exponentially rapidly as $\rho_{0}(t)\sim e^{\sqrt{2t}}$.
(c) For $M>M_{c}$, there is no steady state and the system undergoes
chemotactic collapse. For $M>M_{c}$ the collapse is not exactly
self-similar. In a finite time $t_{coll}$, the system develops a
Dirac peak of mass $M_{c}$ surrounded by a halo whose tail decreases
as $\rho\sim r^{-\alpha(t)}$ with $\alpha(t)$ converging extremely
slowly to $\alpha=2$ for $t\rightarrow t_{coll}$, so that the
evolution looks self-similar. This corresponds to $\rho({\bf
r},t)\rightarrow M_{c}\delta({\bf r})+\tilde \rho({\bf r},t)$. This
non self-similar collapse has been studied in detail in \cite{sc}.
Some results had been obtained earlier by Herrero \& Velazquez
\cite{herrero} but they are different from those of \cite{sc}. The
reason is not well-understood but it may be that the solutions
constructed in \cite{herrero}, while mathematically correct, are
unstable. By contrast, the solution of \cite{sc} shows a very good
agreement with direct numerical simulations of the reduced
Keller-Segel model in $d=2$ (as explained previously, the work
\cite{sc} is presented for self-gravitating Brownian particles but
the results also apply to biological populations). From the Virial
theorem Eq.~(\ref{ks20}), we note that $\dot I\le \epsilon<0$ for
$M>M_{c}$ so that the moment of inertia tends to its minimum value
$I=0$ in a finite time $t_{end}$. This corresponds to the formation
of a Dirac peak at ${\bf r}={\bf 0}$ containing the whole mass $M$.
Therefore, we expect a post-collapse regime for $t_{coll}\le t\le
t_{end}$ leading ultimately to a Dirac peak $\rho({\bf
r})=M\delta({\bf r})$ containing all the mass at $t_{end}$. To our
knowledge, this post-collapse regime has not yet been fully
characterized.

Consider now the case of 2D systems in an infinite domain. (a) For
$M<M_{c}$, $I(t)\rightarrow +\infty$ for $t\rightarrow +\infty$ and
the system evaporates. This regime has been studied in
\cite{virial1}. (b) For $M=M_{c}$, $dI/dt=0$ so that the moment of
inertia is conserved.  It is shown analytically in
\cite{virial1} that the system forms a Dirac peak for $t\rightarrow
+\infty$ and ejects a tiny amount of mass at large distances so as to
satisfy the moment of inertia constraint. In an unbounded domain \cite{virial1}, the central
density diverges logarithmically with time, $\rho_{0}(t)\sim \ln t$,
while the divergence is exponentially fast in a bounded domain
\cite{sc}. For $M=M_{c}$, the system can
also evaporate. (c) For $M>M_{c}$, $I(t)=0$ in a finite time $t_{end}$
so the system collapses to a Dirac peak containing all the mass in a
finite time. The evolution is expected to be similar to the case of a
box-confined system studied in
\cite{sc}.

In $d=3$, in a bounded domain, there exists a critical mass
$M_{c}=31.7 {DR\over \lambda\chi}$ which depends on the box radius
$R$ (contrary to the case $d=2$). This is the equivalent of the
Emden temperature $k_{B}T_{c}={GMm\over 2.52 R}$ in astrophysics.
(a) For $M\le M_{c}$, there exists {\it metastable} equilibrium
states, i.e. local minima of the functional (\ref{chemo2}), to which
the system is attracted for generic initial conditions. (b) For
$M>M_{c}$, there is no equilibrium state and the system undergoes a
self-similar collapse leading to a finite time singularity at
$t=t_{coll}$. The collapse time behaves with the distance to the
critical point as $t_{coll}=t_{*}(\eta-\eta_{c})^{-1/2}$ where
$\eta=\lambda M\chi/4\pi DR$ and $t_{*}=0.91767702...$ \cite{tcoll}.
This pre-collapse regime can be described analytically
\cite{crs,sc}. The density profile decreases as $r^{-2}$. The
central density increases with time as $\rho_{0}\sim
(t_{coll}-t)^{-1}$ and the core radius decreases as $r_{0}(t)\sim
(t_{coll}-t)^{1/2}$.  A Dirac peak is formed in the post-collapse
regime for $t>t_{coll}$. The mass of the Dirac peak increases as
$M_{0}(t)\sim (t-t_{coll})^{1/2}$.  The Dirac peak accrets all the
mass for $t\rightarrow +\infty$ \cite{post}. (c) The peculiar case
$D=0$ (no diffusion) where the particles are driven solely by the
chemotactic drift is equivalent to the case $T=0$ for
self-gravitating Brownian particles described in Sec.
\ref{sec_qualb}.

In an unbounded domain in $d=3$, there is no steady state. The
system can either collapse (as in a bounded domain) or evaporate.
The evaporation process has been treated in \cite{virial1} and the
collapse in \cite{crs,sc}. The choice between collapse or
evaporation probably depends on a complicated notion of basin of
attraction that is function of the initial condition.

Finally, in $d=1$ the problem can be mapped on the Burgers equation
which has the explicit solution (\ref{one7}) in an unbounded domain
and (\ref{dynun3}) in a bounded domain.

\section{Conclusion}

In this paper, we have stressed the analogy between the chemotaxis
of bacterial populations and the dynamics of self-gravitating
Brownian particles. In particular, in $d=2$ dimensions, we have
shown that the critical mass of bacterial populations $M_{c}$ is the
counterpart of the critical temperature $T_{c}$ of self-gravitating
brownian particles. These analogies are not well-known because these
topics (chemotaxis and gravity) are usually studied by very
different communities and the self-gravitating Brownian gas model
has been introduced only recently in physics \cite{crs}. Yet, we
think that the inter-relation between these disciplines is important
to develop and the present paper is a step in that direction. In
particular, we have obtained the value of the critical mass by using
a relation which turns out to be equivalent to the Virial theorem in
astrophysics. We think that many other connections can be made
between the two disciplines, and this will be considered in future
works.

We have also qualitatively discussed the dynamical evolution of a
self-gravitating Brownian gas (or a chemotactic system) in different
dimensions of space by presenting a synthesis of the results obtained
in Chavanis \& Sire
\cite{crs,sc,post,tcoll,banach,crrs,sopik,virial1,virial2}. This gives
a clear picture of the collapse dynamics of a {\it spherically
symmetric} system to a single cluster. If we come back to the general
problem which does not need to be spherically symmetric, we expect
that several collapses will take place at different locations of the
domain (if sufficiently large). Each collapse will be described by the
spherical solution that we have found. But the resulting clusters will
themselves have a non-trivial dynamics and will ``merge'' together so
that their number will decrease with time until a single Dirac peak
containing the whole mass remains at the end. In that case, the
evolution toward the final Dirac peak is progressive. This problem
shares some analogies with the dynamics of vortices in 2D decaying
turbulence
\cite{pomeau,decay} although the equations of motion are of course
different. The analogy with 2D turbulence may be interesting to
develop. The aggregation of clumps in our gravitational Brownian model
\cite{crs} could also be studied by exploiting the analogy with the
Burgers equation  \cite{frisch}. These are directions of
investigation that we plan to explore in the future.

\appendix

\section{A remark on the boundary conditions}
\label{sec_bc}

When we consider the more general Keller-Segel model
\cite{keller,jager,horstmann,crrs}
\begin{equation}
\label{bc1} {\partial\rho\over\partial t}=D\Delta\rho-\chi\nabla (\rho\nabla c),
\end{equation}
\begin{equation}
\label{bc2}{\partial c\over\partial t}=D'\Delta c+a\rho-bc,
\end{equation}
the boundary conditions are the Neumann conditions
\begin{equation}
\label{bc3}
\nabla\rho\cdot {\bf n}=\nabla c\cdot {\bf n}=0,
\end{equation}
where ${\bf n}$ is a unit vector normal to the boundary of the
box. Equation (\ref{bc2}) describes the evolution of the concentration
of the chemical. The chemical diffuses with a diffusion coefficient
$D'$, is created by the bacteria at a rate $a$ and is degraded at a
rate $-b$. With the Neumann boundary conditions, there is no current
of particles (bacteria and chemical) accross the box so we have
$\rho=c=0$ outside the box. From Eqs. (\ref{bc1}) and (\ref{bc3}), we
note that the average concentration of bacteria
$\overline{\rho}(t)=M/V=\overline{\rho}_{0}$ is conserved. From
Eqs. (\ref{bc2}) and (\ref{bc3}), we note that the average
concentration of chemical satisfies
$d\overline{c}/dt+b\overline{c}=a\overline{\rho}$ so that
$\overline{c}(t)=(a\overline{\rho}_{0}/b)(1-e^{-bt})+\overline{c}_{0}e^{-bt}$
tending to $\overline{c}(+\infty)=a\overline{\rho}_{0}/b$ at
equilibrium.  Following J\"ager \& Luckhaus \cite{jager}, we set
$a=\lambda D'$ and consider the limit $D'\rightarrow +\infty$ with
$\lambda\sim 1$. This leads to
\begin{equation}
\label{bc4} {\partial\rho\over\partial t}=D\Delta\rho-\chi\nabla (\rho\nabla c),
\end{equation}
\begin{equation}
\label{bc5}\Delta c=-\lambda (\rho-\overline{\rho}_{0}).
\end{equation}
This model is still well-posed mathematically with the Neumann
boundary conditions (\ref{bc3}). We emphasize that only the {\it
gradient} of concentration $\nabla c$ enters in Eqs.
(\ref{bc3})-(\ref{bc5}). Therefore, the concentration
$c$ itself is un-determined when we make the above-mentioned
approximations since it is obtained only within an additive constant.
When the density blows up (chemotactic collapse) so that $\rho({\bf
r},t)\gg
\overline{\rho}_{0}$, it is justified to consider the model
(\ref{ks1})-(\ref{ks2}) where Eq. (\ref{bc5}) is replaced by a Poisson
equation. It is only in that case (large diffusivity of the chemical
$D'\rightarrow +\infty$ and high concentration $\rho({\bf r},t)\gg
\overline{\rho}_{0}$ of the bacteria) that the Keller-Segel model for
the chemotaxis becomes equivalent to the Smoluchowski-Poisson system
for self-gravitating Brownian particles. Some authors have studied the
model (\ref{ks1})-(\ref{ks2}) on general grounds, i.e. not necessarily
being an approximation valid when $\rho({\bf r},t)\gg
\overline{\rho}_{0}$. We stress, however, that this model  (\ref{ks1})-(\ref{ks2}) is
not well-posed mathematically with the Neumann boundary conditions
(\ref{bc3}). Indeed, integrating Eq. (\ref{ks2}) and using the
divergence theorem, we have $\oint \nabla c\cdot {\bf n}\ dS=-\lambda
M\neq 0$, so that the Neumann boundary conditions (\ref{bc3}) cannot
be satisfied in that case. One way to circumvent this difficulty is to
use the boundary conditions defined in Sec. \ref{sec_ks}. We must keep
in mind, however, that these boundary conditions do not determine the
physical concentration $c_{phys}({\bf r},t)$ but only a ``field''
$c({\bf r},t)$ that has the same gradient (this field can take
positive or negative values). Furthermore, this field does not satisfy
the requirement $\nabla c\cdot {\bf n}=0$ so it is not clear how close
it is related to the solution of the problem (\ref{bc3})-(\ref{bc5}).
Finally, although the problem (\ref{ks1})-(\ref{ks2}) is well-posed
mathematically with the boundary conditions of Sec. \ref{sec_ks}, we
now find that $c({\bf r},t)\neq 0$ outside the box. This is a physical
problem because there is no reason why the chemical should exit the
material box (unless it has porous properties). Of course, the
practical solution is to consider the concentration $c({\bf r},t)$
inside the box only, and ignore that field outside. Another
possibility is to impose $c=0$ on the boundary of the domain
(Dirichlet) and $c({\bf r},t)=0$ outside. In the case of
self-gravitating Brownian particles, the concentration $-c({\bf r},t)$
is replaced by the gravitational potential $\Phi({\bf r},t)$. We can
have $\Phi({\bf r},t)\neq 0$ outside the box enclosing the particles
because the gravitational field ``traverses'' the box (a material body
enclosed within a container creates a gravitational force outside this
container) so that the boundary conditions defined in
Sec. \ref{sec_sp} are natural. We conclude therefore that the reduced
chemotactic model (\ref{ks1})-(\ref{ks2}) is not very well-posed
physically compared with the initial problem
(\ref{bc1})-(\ref{bc3}). By contrast, the Smoluchowski-Poisson system
(\ref{sp1})-(\ref{sp2}) is rigorous on a physical point of view as the
gravitational potential is always solution of the Newton-Poisson
equation (\ref{sp2}) with the boundary conditions given in
Sec. \ref{sec_sp}.

\section{Overdamped Virial theorem for spherically symmetric systems}
\label{sec_vtsss}

In this Appendix, we show that the overdamped Virial theorem
(\ref{sp4}) can be established very simply in the case of spherically
symmetric systems. In that case, the moment of inertia can be
expressed as a function of the mass profile in the form
\begin{equation}
I(t)=\int_{0}^{R}\frac{\partial M}{\partial r}(r,t)\; r^{2}dr=MR^{2}-2\int_{0}^{R}M(r,t)r\; dr,
\label{vtsss1}
\end{equation}
where the second equality follows from an integration by parts. Taking the time derivative of Eq. (\ref{vtsss1}), inserting Eq. (\ref{one9}) and using simple integrations by parts, we obtain
\begin{equation}
\frac{1}{2}\xi \dot I=-\frac{k_{B}T}{m}\lbrack RM'(R,t)-dM\rbrack -\int_{0}^{R}\frac{GM(r,t)}{r^{d-2}}\frac{\partial M}{\partial r}(r,t)\; dr.
\label{vtsss2}
\end{equation}
This can be rewritten in the form
\begin{equation}
\frac{1}{2}\xi \dot I=dNk_{B}T -\int_{0}^{R}\frac{GM(r,t)}{r^{d-2}}\frac{\partial M}{\partial r}(r,t)\; dr -dPV,
\label{vtsss3}
\end{equation}
where we have used $P=k_{B}T\rho(R)/m$ and $V=\frac{1}{d}S_{d}R^{d}$. Noting finally  (see the Appendix of \cite{virial1}) that 
\begin{equation}
W_{ii}=-\int_{0}^{R}\frac{GM(r,t)}{r^{d-2}}\frac{\partial M}{\partial r}(r,t)\; dr,
\label{vtsss4}
\end{equation}
we obtain Eq. (\ref{sp4}). In particular, for $d=2$ we immediately deduce
Eq. (\ref{sp6}) from Eq. (\ref{vtsss3}).

\section{Explicit solution of the 2D Boltzmann-Poisson equation}
\label{sec_bp}

In this Appendix, we explicitly solve the Boltzmann-Poisson equation
(\ref{chemo3}) in $d=2$ for axisymmetric solutions. The density (\ref{chemo1}) can
be rewritten
\begin{equation}
\rho=\rho_{0}e^{\frac{\chi}{D}(c-c_{0})},
\label{bp1}
\end{equation}
where $\rho_{0}$ and $c_{0}$ are the values of the density at the
center of the domain. The Boltzmann-Poisson equation takes the form
\begin{equation}
{1\over r}{d\over dr}\left (r{dc\over dr}\right )=-\lambda \rho_{0}e^{\frac{\chi}{D}(c-c_{0})}.
\label{bp2}
\end{equation}
Introducing $\psi=-\frac{\chi}{D}(c-c_{0})$ and $\xi=(\lambda\chi\rho_{0}/D)^{1/2}r$, we obtain
\begin{equation}
{1\over \xi}{d\over d\xi}\left (\xi{d\psi\over d\xi}\right )=e^{-\psi},
\label{bp3}
\end{equation}
\begin{equation}
\psi(0)=\psi'(0)=0.
\label{bp4}
\end{equation}
With the change of variables
$t=\ln\xi$ and $\psi=2\ln\xi-z$, Eq. (\ref{bp3}) can be rewritten
\begin{equation}
{d^{2}z\over dt^{2}}=-e^{z}=-{d\over dz}(e^{z}).
\label{bp5}
\end{equation}
This corresponds to the motion of a  particle in a potential $V(z)=e^{z}$. Using the initial condition (\ref{bp4}) which translates into ($z\rightarrow -\infty$, $dz/dt=2$) for $t\rightarrow -\infty$, the first integral is
\begin{equation}
\frac{1}{2}\left (\frac{dz}{dt}\right )^{2}+e^{z}=2.
\label{bp6}
\end{equation}
This first order differential equation is readily integrated yielding
\begin{equation}
\tanh^{-1}\sqrt{1-\frac{1}{2}e^{z}}=t+C.
\label{bp7}
\end{equation}
Returning to original variables, we get
\begin{equation}
e^{-\psi}={8\lambda^{2}\over  (1+\lambda^{2}\xi^{2} )^{2}},
\label{bp8}
\end{equation}
where $\lambda$ is a constant of integration related to $C$. It is determined by $\psi(0)=0$ yielding $8\lambda^{2}=1$ so we finally obtain
\begin{equation}
e^{-\psi}={1\over  (1+{1\over 8}\xi^{2} )^{2}}.
\label{bp9}
\end{equation}
Using the Gauss theorem, the total mass is given by
\begin{equation}
\frac{dc}{dr}(R)=-\frac{\lambda M}{2\pi R}.
\label{bp10}
\end{equation}
Let $\alpha=(\lambda\chi\rho_{0}/D)^{1/2}R$ denote the value of $\xi$ at the edge of the box so that $\xi=\alpha r/R$. With these notations, the foregoing relation can be rewritten
\begin{equation}
\alpha\psi'(\alpha)=\frac{\lambda M\chi}{2\pi D}.
\label{bp11}
\end{equation}
Using Eq. (\ref{bp8}) and introducing the critical mass (\ref{chemo7}), we find that this relation is equivalent to
\begin{equation}
\frac{\alpha^{2}}{8}=\frac{M}{M_{c}-M}.
\label{bp12}
\end{equation}
Recalling the definition of $\alpha$, this equation determines the relation between the mass and the central density according to
\begin{equation}
\rho_{0}=\frac{M_{c}}{\pi R^{2}}\frac{M/M_{c}}{1-\frac{M}{M_{c}}}.
\label{bp13}
\end{equation}
Finally, noting that
\begin{equation}
\rho={\rho_{0}\over  (1+{\alpha^{2}\over 8}(\frac{r}{R})^{2} )^{2}},
\label{bp14}
\end{equation}
and using Eqs. (\ref{bp12}) and (\ref{bp13}) we finally obtain
Eq. (\ref{chemo10}). On the other hand, in an infinite domain, the
Gauss theorem $\lim_{\xi\rightarrow
+\infty}\xi\psi'(\xi)=\frac{\lambda M\chi}{2\pi D}$ and
Eq. (\ref{bp9}) imply that $M=M_{c}=\frac{8\pi D}{\chi\lambda}$. Then,
using Eqs. (\ref{bp1}) and (\ref{bp9}) we obtain Eq. (\ref{chemo12}).

The two-dimensional Boltzmann-Poisson equation (\ref{chemo3}) appeared
in very different topics: self-gravitating isothermal gaseous
cylinders in hydrostatic equilibrium \cite{ostriker,stodolkiewicz},
statistical equilibrium states of two-dimensional stellar systems in
the microcanonical ensemble \cite{klb,ap}, statistical equilibrium
states of two-dimensional self-gravitating Brownian particles in the
canonical ensemble \cite{sc}, statistical mechanics of point vortices
in two-dimensional hydrodynamics \cite{jm,lp,caglioti,Chouches},
chemotaxis of bacterial populations \cite{herrero}.

\section{Explicit solution of the 1D Boltzmann-Poisson equation}
\label{sec_bpun}

In this Appendix, we explicitly solve the Boltzmann-Poisson equation
(\ref{chemo3}) in $d=1$. Writing $\rho=\rho_0 e^{-\psi}$ with
$\psi=\beta m (\Phi-\Phi_0)$ where $\rho_0$ is the central density
and $\Phi_0$ the central potential, and introducing the scaled
distance $\xi=(2G\beta m\rho_0)^{1/2}x$, the Boltzmann-Poisson
equation in $d=1$ can be written
\begin{equation}
\frac{d^2\psi}{d\xi^2}=e^{-\psi},\label{exunbp1}
\end{equation}
with $\psi(0)=\psi'(0)=0$. This is similar to the equation of motion
of a particle in a potential $V(\psi)=e^{-\psi}$. The first integral
is $E=\frac{1}{2}(d\psi/d\xi)^2+e^{-\psi}$. Using the initial
conditions we find that $E=1$. Therefore, we get
\begin{equation}
\frac{d\psi}{\sqrt{2(1-e^{-\psi})}}=d\xi.
\label{exunbp2}
\end{equation}
This is integrated into
\begin{equation}
\tanh^{-1}\sqrt{1-e^{-\psi}}=\frac{1}{\sqrt{2}}\xi, \label{exunbp3}
\end{equation}
and we finally obtain
\begin{equation}
e^{-\psi}=\frac{1}{\cosh^{2}(\xi/\sqrt{2})}. \label{exunbp4}
\end{equation}
In a infinite domain, the Gauss theorem $(d\Phi/dx)(+\infty)=GM$
in scaled variables becomes
\begin{equation}
\lim_{\xi\rightarrow +\infty}\psi'(\xi)=M\left (\frac{\beta G
m}{2\rho_0}\right )^{1/2}. \label{exunbp5}
\end{equation}
Using Eq. (\ref{exunbp4}), this yields Eq. (\ref{one3})-b. Then, Eq. (\ref{exunbp4}) yields Eq. (\ref{one2}). In a finite domain, we call $\alpha=(2\beta G m\rho_0)^{1/2}R$
the value of the scaled distance $\xi$ at the box radius $R$. Then,
we have $\xi=(\alpha/R)x$. The Gauss theorem $(d\Phi/dx)(R)=GM$ in
scaled variables becomes
\begin{equation}
\alpha\psi'(\alpha)=\beta GMmR\equiv \eta. \label{exunbp6}
\end{equation}
Using Eq. (\ref{exunbp4}), we find that
\begin{equation}
\eta=\sqrt{2}\alpha \tanh(\alpha/\sqrt{2}). \label{exunbp7}
\end{equation}
We also have by definition
\begin{equation}
\frac{2R\rho_{0}}{M}=\frac{\alpha^{2}}{\eta}. \label{exunbp7n}
\end{equation}
Therefore, Eqs. (\ref{exunbp7}) and (\ref{exunbp7n}) determine the central density in terms of the temperature. Finally, the density and mass profiles can be written
\begin{equation}
\rho(x)=\frac{\rho_0}{\cosh^{2}(\alpha x/\sqrt{2}R)}, \qquad M(x)=\frac{M}{\tanh(\alpha/\sqrt{2})}\tanh(\alpha x/\sqrt{2}R).
\label{exunbp8}
\end{equation}
The central density monotonically increases as the temperature
decreases (see Fig. \ref{etarhod1}). At $T=0$, the density profile is
a Dirac peak: $\rho(x)=M\delta(x)$ (see Fig. \ref{rhoxd1}). Noting
that $\alpha=\beta GmRa/\sqrt{2}$, these results are equivalent to
those of Sec. \ref{sec_onefin}. The one-dimensional
Boltzmann-Poisson equation (\ref{exunbp1}) appeared in very different
topics: highly flattened galactic disks
\cite{spitzer}, stellar systems stratified in plane parallel layers
\cite{camm}, one dimensional self-gravitating Brownian particles
\cite{sc}, statistical mechanics of two-dimensional turbulence in
a shear layer \cite{sommeria}.

% \appendix

% \section{}
% \label{}

% Bibliographic references with the natbib package:
% Parenthetical: \citep{Bai92} produces (Bailyn 1992).
% Textual: \citet{Bai95} produces Bailyn et al. (1995).
% An affix and part of a reference:
%   \citep[e.g.][Ch. 2]{Bar76}
%   produces (e.g. Barnes et al. 1976, Ch. 2).


\begin{thebibliography}{}

% \bibitem[Names(Year)]{label} or \bibitem[Names(Year)Long names]{label}.
% (\harvarditem{Name}{Year}{label} is also supported.)

\bibitem{houches}  {\small {\it Dynamics and thermodynamics of systems with long range interactions}, edited by Dauxois, T., Ruffo, S., Arimondo, E. and  Wilkens, M. Lecture Notes in Physics, Springer (2002)}

\bibitem{murray}  {\small J.D. Murray, {\it Mathematical Biology} (Springer, Berlin, 1991).}

\bibitem{keller}  {\small E. Keller, L.A. Segel J. theor. Biol. {\bf 26}, 399 (1970).}

\bibitem{jager}  {\small W. J\"ager, S. Luckhaus, Trans. Amer. Math. Soc.  {\bf 329}, 819 (1992).}

\bibitem{horstmann}  {\small D. Horstmann, Jahresberichte der DMV  {\bf 106}, 51 (2004).}

\bibitem{crs}  {\small P.H. Chavanis, C. Rosier and C. Sire, Phys. Rev. E {\bf  66}, 036105 (2002).}

\bibitem{sc}  {\small C. Sire and P.H. Chavanis, Phys. Rev. E {\bf 66}, 046133 (2002).}

\bibitem{post}  {\small C. Sire and P.H. Chavanis, Phys. Rev. E {\bf 69}, 066109 (2004).}

\bibitem{tcoll}  {\small P.H. Chavanis and C. Sire, Phys. Rev. E {\bf 70}, 026115 (2004).}

\bibitem{banach}  {\small  C. Sire and P.H. Chavanis, Banach Center Publ. {\bf 66}, 287 (2004).   }

\bibitem{crrs}  {\small  P.H. Chavanis, M. Ribot, C. Rosier  and C. Sire, Banach Center Publ. {\bf 66}, 103 (2004).   }

\bibitem{sopik}  {\small J. Sopik, C. Sire and P.H. Chavanis, Phys. Rev. E {\bf 72}, 026105 (2005).}

\bibitem{virial1}  {\small P.H. Chavanis and C. Sire, Phys. Rev. E {\bf 73}, 066103 (2006). }

\bibitem{virial2}  {\small P.H. Chavanis and C. Sire, Phys. Rev. E {\bf 73}, 066104 (2006). }

\bibitem{cp}  {\small S. Childress, J.K. Percus, Math. Biosci. {\bf 56}, 217 (1981).}

\bibitem{childress}  {\small S. Childress, Lecture Notes in Biomath. {\bf 55}, 61 (1984).}

\bibitem{nagai}  {\small T. Nagai, Adv. Math. Sci. Appl.  {\bf 5}, 581 (1995).}

\bibitem{herrero}  {\small M.A. Herrero, J.J.L. Velazquez, Math. Ann.  {\bf 306}, 583 (1996).}

\bibitem{dolbeault}  {\small J. Dolbeault, B. Perthame, C. R. Acad. Sci. Paris, Ser. I  {\bf 339}, 611 (2004).}

\bibitem{biler1}  {\small P. Biler, G. Karch, P. Lauren\c cot, T. Nadzieja, Topol. Methods Nonlinear Anal. {\bf 27}, 133 (2006).}

\bibitem{biler2}  {\small P. Biler, G. Karch, P. Lauren\c cot, T. Nadzieja,  Math. Methods Appl. Sci.  {\bf 29}, 1563  (2006).}

\bibitem{lang}  {\small  P.H. Chavanis and C. Sire, Phys. Rev. E {\bf 69}, 016116 (2004). }

\bibitem{logo}  {\small  P.H. Chavanis and C. Sire, Physica A {\bf 375}, 140 (2007). }

\bibitem{hb}  {\small  P.H. Chavanis, Physica A {\bf 361}, 55 (2006);  P.H. Chavanis, Physica A {\bf 361}, 81 (2006). }

\bibitem{salsberg}  {\small A.M. Salsberg, J. Math. Phys. {\bf 6}, 158 (1965).}

\bibitem{klb}  {\small J. Katz, D. Lynden-Bell, Mon. Not. R. Astron. Soc. {\bf 184}, 709 (1978).}

\bibitem{paddy}  {\small T. Padmanabhan, Phys. Rep.  {\bf 188}, 287 (1990).}

\bibitem{new}  {\small P.H. Chavanis, [cond-mat/0612124] }

\bibitem{ostriker}  {\small J. Ostriker, ApJ {\bf 140}, 1056 (1964).}

\bibitem{stodolkiewicz}  {\small J.S. Stodolkiewicz, Acta Astr. {\bf 13}, 30 (1963).}

\bibitem{ap}  {\small J.J Aly, J. Perez, Phys. Rev. E {\bf 60}, 5185 (1999).}

\bibitem{at}  {\small E. Abdalla, M.R. R Tabar Phys. Lett. B  {\bf 440}, 339 (1998).}

\bibitem{lp}  {\small T.S. Lundgren, Y.B. Pointin, J. Stat. Phys. {\bf 17}, 323 (1977).}

\bibitem{caglioti}  {\small E. Caglioti, P.L. Lions, C. Marchioro, M. Pulvirenti, Commun. Math. Phys. {\bf 143}, 501 (1992).}

\bibitem{Chouches}  {\small P.H. Chavanis, in {\it Dynamics and thermodynamics of systems with long range interactions}, edited by Dauxois, T., Ruffo, S., Arimondo, E. and  Wilkens, M. Lecture Notes in Physics, Springer (2002); see [cond-mat/0212223]  }


\bibitem{degrad}  {\small P.H. Chavanis, Eur. Phys. J. B {\bf 54}, 525 (2006). }

\bibitem{gen}  {\small P.H. Chavanis, Phys. Rev. E {\bf 68}, 036108 (2003).}

\bibitem{ijmpb}  {\small P.H. Chavanis,  Int J. Mod. Phys. B {\bf 20}, 3113 (2006).}

\bibitem{chandra}  {\small S. Chandrasekhar, {\it An Introduction to the Theory of Stellar Structure} (Dover, 1942).}


\bibitem{burgers}  {\small J. Burgers, {\it The Nonlinear Diffusion Equation} (D. Reidel, Publ. Co., 1974).}

\bibitem{frisch}  {\small M. Vergassola, B. Dubrulle, U. Frisch and A. Noullez, Astron. Astrophys.   {\bf 289}, 325 (1994).}

\bibitem{hopf}  {\small E. Hopf, Comm. Pure App. Mech    {\bf 3}, 201 (1950).}

\bibitem{cole}  {\small J. Cole, Quart. Appl. Math.   {\bf 9}, 225 (1951).}

\bibitem{acedo}  {\small L. Acedo, Europhysics Letters  {\bf 73}, 5 (2006).}

\bibitem{pomeau}{\small  G.F. Carnevale, J.C. McWilliams, Y. Pomeau, J.B. Weiss and W.R. Young, Phys. Rev. Lett. {\bf 66}, 2735 (1991).}

\bibitem{decay}{\small  C. Sire, P.H. Chavanis, Phys. Rev. E {\bf 61}, 6644 (2000).}

\bibitem{jm}{\small  G. Joyce \& D. Montgomery, J. Plasma Phys. {\bf 10}, 107 (1973).}

\bibitem{spitzer}  {\small L. Spitzer, Astrophys. J.  {\bf 95}, 329 (1942).}

\bibitem{camm}  {\small G.L. Camm, Mon. Not. R. Astron. Soc.   {\bf 110}, 305 (1950).}

\bibitem{sommeria}  {\small J. Sommeria, C. Staquet and R. Robert, J. Fluid Mech.   {\bf 233}, 661 (1991).}













\end{thebibliography}
\end{document}